# Novel Field-Induced Phases in HoMnO$_3$ at Low Temperatures

B. Lorenz[1], F. Yen[1], M. M. Gospodinov[2], and C. W. Chu[1,3,4]

[1]*Department of Physics and TCSAM, University of Houston, Houston, TX 77204-5002*

[2]*Institute of Solid State Physics, Bulgarian Academy of Sciences, 1784 Sofia, Bulgaria*

[3]*Lawrence Berkeley National Laboratory, 1 Cyclotron Road, Berkeley, CA 94720*

[4]*Hong Kong University of Science and Technology, Hong Kong, China*

Abstract

The novel field-induced re-entrant phase in multiferroic hexagonal HoMnO$_3$ is investigated to lower temperatures by dc magnetization, ac susceptibility, and specific heat measurements at various magnetic fields. Two new phases have been unambiguously identified below the Néel transition temperature, $T_N$=76 K, for magnetic fields up to 50 kOe. The existence of an intermediate phase between the P$\underline{6_3}$$\underline{c}$m and P$\underline{6_3}$$\underline{cm}$ magnetic structures (previously predicted from dielectric measurements) was confirmed and the magnetic properties of this phase have been investigated. At low temperatures (T<5 K) a dome shaped phase boundary characterized by a magnetization jump and a narrow heat capacity peak was detected between the magnetic fields of 5 kOe and 18 kOe. The transition across this phase boundary is of first order and the magnetization and entropy jumps obey the magnetic analogue of the Clausius-Clapeyron relation. Four of the five low-temperature phases coexist at a tetracritical point at 2 K and 18 kOe. The complex magnetic phase diagram so derived provides an informative basis for unraveling the underlying driving forces for the occurrence of the various phases and the coupling between the different orders.

# I. INTRODUCTION

The coexistence and mutual interference of different types of long-range orders, such as (anti) ferromagnetic, (anti) ferroelastic, and ferroelectric, have long inspired researchers because of their fundamental interest and their significance for potential applications. Rare-earth hexagonal manganites, RE-MnO$_3$ (RE=Sc, Y, Er, Ho, Tm, Yb, Lu) have attracted increasing attention since the discovery of a dielectric anomaly in YMnO$_3$ at the antiferromagnetic (AFM) ordering temperature of the Mn$^{3+}$ magnetic moments, $T_N \approx 80$ K.[1] The hexagonal rare-earth manganites exhibit a ferroelectric (FE) order at high Curie temperatures ($T_C$) between 590 and 1000 K and an AFM order of manganese spins, at low Neel temperatures ($T_N$) < 100 K, clearly showing the coexistence of the FE and AFM orders. In the hexagonal structure (space group P6$_3$cm) the Mn$^{3+}$ form triangular planar sublattices stacked along the c-axis and neighboring spins are coupled antiferromagnetically via the oxygen ions by superexchange interaction. The AFM exchange coupling in a triangular lattice gives rise to spin frustration effects and, at $T_N$ the Mn$^{3+}$ moments order in a way so that neighboring Mn-moments form a 120° angle.[2] In addition, most rare-earth ions (e.g. Ho$^{3+}$ but not Y$^{3+}$) carry their own magnetic moment oriented along the c-axis of the P6$_3$cm structure. The rare-earth moment can interact with the Mn$^{3+}$ spins and the dielectric polarization and thus increase the complexity of the phase diagram and the physical phenomena that can be observed. For example, the magnetic phase diagram of HoMnO$_3$ studied by neutron scattering [3,4] and second harmonic generation optical experiments [5,6] shows two additional phase transitions below $T_N$ indicating subtle changes in the magnetic order of the Mn$^{3+}$ and Ho$^{3+}$ ions at zero external magnetic field. At $T_{SR} \approx 33$ K a sharp Mn-spin reorientation transition takes place at which all Mn-moments rotate in-plane by an angle of 90° changing the magnetic symmetry from P$\underline{6}_3$cm (T>$T_{SR}$) to P$\underline{6}_3$c$\underline{m}$ (T<$T_{SR}$). At even lower temperatures, $T_2 \approx 5$ K, another change of the magnetic structure was reported but the magnetic order in this phase is still a matter of discussion.[4,6] The transitions at $T_{SR}$ and $T_2$ are accompanied by partial or complete magnetic ordering of the Ho$^{3+}$ moments but the detail of the Ho-spin order has not been resolved yet. All magnetic transitions are well below the FE Curie temperature of $T_C$=830 K.

Dielectric anomalies at the magnetic transitions in RE-MnO$_3$ have first been observed in YMnO$_3$ at the Néel temperature [1] and were later found at $T_N$ in other hexagonal RE-MnO$_3$ compounds.[7,8,9], suggesting the coupling between the magnetic and ferroelectric orders. They are of particular physical interest because, by symmetry arguments, a direct coupling between the in-plane staggered magnetization and the c-axis polarization in the P6$_3$cm structure is not allowed.[10] Indirect coupling between magnetization and polarization via lattice strain or other secondary effects have been proposed, as evidenced by later observations.[11] Very recently we have observed a sharp and distinct peak of the dielectric constant at the spin-reorientation transition of HoMnO$_3$ that is indicative of an unusual magneto-dielectric coupling of the different orders coexisting in this compound.[11] The dielectric peak evolves into a plateau-like structure in external magnetic fields, H, that led us to propose the existence of a new re-entrant novel phase (INT phase) separating the P$\underline{6}_3$cm and P$\underline{6}_3$c$\underline{m}$ phases in the T-H phase space. The dielectric anomalies have been found to be accompanied by small but distinct changes of the c-axis magnetization.[11] The magnetic phase diagram of $HoMnO_3$ is apparently far more complex than previously reported.[6]

We have measured systematically the dc magnetization, ac magnetic susceptibility, and specific heat of $HoMnO_3$ for temperatures T down to 1.8 K and magnetic fields H up to 50 kOe. The novel INT phase, recently proposed by us based on dielectric measurements above 5 K,[11] is found to evolve into other phases at lower temperatures and at magnetic fields > 5 kOe. A new, dome shaped phase boundary was detected at T < 3.3 K in 5 kOe < H < 18 kOe, characterized by a sharp change of the c-axis magnetization and a narrow peak of the heat capacity. The observations define a complex phase diagram of $HoMnO_3$ and provide an informative basis for further exploration of the occurrence of the various orders and interactions between them.

## II. EXPERIMENTAL

Pure polycrystalline hexagonal $HoMnO_3$ was synthesized by a solid-state reaction of stoichiometric amounts of $Ho_2O_3$ (99.99 %) and $MnO_2$ (99.99 %), and further annealed for 24h at 1120 °C in oxygen atmosphere. $HoMnO_3$ single crystals were grown by the High Temperature Solution Growth Method using $PbF_2/PbO/B_2O_3$ flux ($PbF_2$ : PbO : $B_2O_3$ = 0.8 : 0.195 : 0.005). The flux was mixed with $HoMnO_3$ powder in a 7 : 1 ratio and annealed in a platinum crucible at 1250 °C for 48h in oxygen. After that the temperature was decreased to 1000 °C at a rate of 0.5 °C/h. The flux was decanted and well-shaped hexagonal plate-like crystals of typical size 3 x 5 x 0.2 $mm^3$ were removed from the bottom of the crucible. Fig. 1 shows an example of the single crystals used for magnetic measurements. The dimensions of this crystal are 2.6 x 1.6 x 0.26 $mm^3$.

The dc magnetization was measured employing the Magnetic Property Measurement System (Quantum Design). The ac susceptibility experiments were conducted using the same system at a frequency of 355 Hz and an ac field amplitude of 3 Oe in an external dc magnetic field of up to 50 kOe. The magnetic field was oriented parallel as well as perpendicular to the hexagonal c-axis. Data have been acquired as a function of temperature and as a function of external field. The heat capacity was measured as a function of temperature above 1.8 K and in magnetic field parallel to the c-axis using the Physical Property Measurement System (Quantum Design). The typical sample mass for this measurement was about 3 mg. Special care had been taken in all high field experiments to avoid the movement or rotation of the single crystal caused by a field induced magnetic moment in the a-b plane (not aligned with the external field) and the resulting torque.

## III. RESULTS AND DISCUSSION

### A. Magnetic Measurements

The temperature dependence of the inverse magnetization, $\chi_{dc}^{-1}=H/M$, with the magnetic field oriented along the c- and a-axis is shown in Fig. 2 over a large temperature range. H is the magnetic field experienced by the sample after applying the demagnetization correction. Above about 100 K, $\chi_{dc}(T)$ follows the Curie-Weiss law for magnetic fields oriented in-plane as well as along the c-axis. The estimated effective moment, $\mu_{eff}$ = 11.4 $\mu_B$, is in good agreement with the data reported for polycrystalline samples [3] and the theoretically expected value of 11.7 $\mu_B$. The extrapolation of the c-axis susceptibility above 100 K reveals the AFM spin coupling along this axis with a Curie-Weiss temperature of $\theta$ = -115 K. This value of $\theta$ is comparable with the results of Ref. [9], but it is distinctly larger than the –17 K estimated from the magnetic data of the polycrystalline samples.[3] The discrepancy can be resolved when we consider the effect of random orientation of grains in polycrystalline materials on the magnetic susceptibility. With H // a-axis the extrapolated Curie temperature equals zero within the error limits of +/- 1 K (Fig. 2), i.e. the sample appears to be paramagnetic with no indication of an exchange coupling between neighboring $Mn^{3+}$ or $Ho^{3+}$ spins. This is a direct consequence of the geometric frustration since both, Mn and Ho ions, form triangular sublattices perpendicular to the c-axis. Although an AFM superexchange interaction leading to the Néel transition at 76 K clearly exists the frustration effects result in a paramagnetic response of the high-temperature susceptibility with respect to an in-plane magnetic field and a zero Curie-Weiss temperature. Averaging our data of the in-plane and c-axis magnetization with the weight factors of 2/3 and 1/3, respectively, and plotting the inverse susceptibility after demagnetization corrections we get a high-T Curie-Weiss like susceptibility with $\theta \approx$ -15 K, close to the value estimated in Ref. [3]. Therefore, the low value of $\theta$ derived from polycrystalline samples is an artifact due to the anisotropy of the magnetic response parallel and perpendicular to the hexagonal c-axis.

Our data show that the magnetic phase diagram of $HoMnO_3$ in external fields is very complex and it is convenient to discuss the high temperature (T > 5 K) and the low temperature regions separately. The temperature dependence of the dc magnetization, M(T), of $HoMnO_3$ below the Néel temperature associated with the ordering of the $Mn^{3+}$ moments at $T_N$ = 76 K, is dominated by the paramagnetic contribution of the $Ho^{3+}$ magnetic moments. M(T) increases strongly with decreasing T. No anomaly could be detected at $T_N$, in agreement with recent reports.[3,11] At lower temperatures, however, two distinct anomalies of the c-axis magnetization are unambiguously shown to exist. The first anomaly appears at 32.8 K, the temperature of the Mn-spin rotation transition, it is indicated by a small, but sharp drop of M(T). The second anomaly at $T_2$ = 5.2 K results in a sharp increase of M(T) followed by a smooth decrease to lower T. Both anomalies are more clearly seen in the inverse magnetization, H/M, shown in Fig. 3 (open symbols, left scale). The sharpness of the two magnetic transitions is obvious from the two narrow peaks that appear in the derivative, d(H/M)/dT, also displayed in Fig. 3 (closed symbols, right scale). The in-plane magnetization shows an analogous increase at $T_2$ (inset of Fig. 2) but no anomaly has been observed at $T_{SR}$. The abrupt change at $T_{SR}$ of the c-axis magnetization has to be related to magnetic order of the $Ho^{3+}$ magnetic moments oriented along the c-axis since the Mn-spins are strictly confined in the a-b plane and their in-plane rotation cannot change the c-axis magnetization. From the decrease of M(T) at $T_{SR}$ we infer that at least part of the Ho-spins order antiferromagnetically and this order is correlated with the Mn-spin rotation transition. Between $T_2$ and $T_{SR}$ the inverse c-axis susceptibility is a linear function of temperature (Fig. 3) with the extrapolated Curie-Weiss parameters $\theta$ = -10 K and $\mu_{eff}$ = 7.6 $\mu_B$. Since all of the $Mn^{3+}$ spins are ordered below 76 K the paramagnetic contribution to the c-axis susceptibility below $T_{SR}$ must be due to the $Ho^{3+}$ moments. The effective moment of 7.6 $\mu_B$ is distinctly lower than the theoretical value of 10.6 $\mu_B$ (expected for paramagnetic $Ho^{3+}$ moments with orbital momentum of J=8). This lends further support for the partial, but not complete, AFM ordering of the Ho moments at the Mn spin reorientation transition.

The AFM order of Ho moments parallel to the c-axis is also supported by the results of neutron scattering experiments,[3,4] although the available data differ with respect to the precise ordering temperature and the proposed magnetic structure. Munoz et al. [3] concluded that the AFM order of 2/3 of the Ho-spins sets in at 25.4 K but Lonkai et al. [4] found the onset of Ho-spin ordering at 32.5 K, exactly the temperature at which we have observed the first anomaly in the c-axis magnetization (Fig. 3). Our recent investigation showing a sharp peak in the c-axis dielectric constant at 32.8 K [11] and the existence of a specific heat anomaly at this temperature (discussed in Section B) lend further support to the idea that the Mn-spin rotation transition does not only modify the in-plane magnetic order of $Mn^{3+}$ but also affects other physical properties (magnetization, dielectric constant) measured along the c-axis. The different subsystems, Mn and Ho moments, dielectric polarization, are obviously correlated and respond in a concerted manner to the change of magnetic order at $T_{SR}$ revealing an interesting coupling between magnetic and dielectric properties.

The effect of external magnetic fields parallel to the c-axis on the magnetic phase transitions is best visualized in Fig. 4 by plotting the derivative d(H/M)/dT. The different curves are offset by a constant for better clarity. It is remarkable that d(H/M)/dT is constant in a broad temperature interval between the spin rotation transition temperature and $T_2$ (curve A and B in Fig. 4) indicating that the Curie-Weiss like spin correlation in this temperature range is barely affected by the magnetic field. Let us tentatively denote the two high temperature magnetic phases separated by the zero-field spin rotation transition temperature $T_{SR}$ as HT1 and HT2 phase, respectively. At higher magnetic fields (H > 20 kOe) the two narrow peaks of d(H/M)/dT develop into two broad features. For each broad anomaly of d(H/M)/dT two characteristic temperatures, $T_1(H)$ and $T_2(H)$, are defined by the deviation of d(H/M)/dT from the constant value, as indicated in Fig. 4 (curve C). At the spin rotation transition, both critical temperatures coincide at zero magnetic field, i.e $T_1(H=0)=T_2(H=0)=T_{SR}$. With increasing magnetic field the two broad features in d(H/M)/dT move towards each other and the temperature range of the HT2 phase (still defined by the constant derivative d(H/M)/dT) shrinks until it disappears completely at about 34 kOe. This reentrant behavior is also observed for the HT1 phase between 20 kOe and 40 kOe. The phase transition lines defined by $T_1(H)$ and $T_2(H)$ in open circles extracted from d(H/M)/dT are given in Fig. 5. Note that there is an intermediate phase (tentatively called INT phase) between $T_1$ and $T_2$ that separates the HT1 and HT2 phases. The existence of this INT phase is in perfect agreement with our recent conclusions based on the observation of the field-affected dielectric anomalies in $HoMnO_3$.[11] From the present magnetic measurements extending to temperatures as low as 1.8 K we conclude that the INT phase extends to very low T in the low field region of the phase diagram and it transforms into other low temperature phases at fields > 5 kOe (note that the previous dielectric investigation was limited to T>5 K). The slope of H/M in the INT phase is steadily increasing with decreasing temperature as indicated by the dotted line in Fig. 6, in which all available data for different magnetic fields are plotted on the same scale.

To further examine the magnetic field effects on the INT phase, we have measured the isothermal ac susceptibility as a function of a dc external field, $\chi_{ac}(H)$. Both fields, the external dc H and the ac modulation field, were directed along the hexagonal c-axis. Fig. 7 shows the real part, $\chi'_{ac}(H)$, for temperatures above 5 K. $\chi'_{ac}(H)$ undergoes a sharp rise at $H_2$ (corresponding to phase boundary $T_2(H)$), followed by a rapid drop at $H_1$ (corresponding to phase boundary $T_1(H)$) as H increases (Fig. 7). The phase boundaries defined by $H_2$ and $H_1$ derived from $\chi'_{ac}(H)$ in Fig. 7 are plotted as solid circles in the phase diagram (Fig. 5) and they are in perfect agreement with the data from dc magnetization measurements. The current results show that the INT phase is characterized by distinct magnetic properties with an enhanced $\chi'_{ac}(H)$ in analogy to the enhanced dielectric constant ε previously reported.[11] The results show that the INT phase is the stable phase below $T_2$ = 5.2 K at zero magnetic field. This explains the step-like increase of the dielectric constant observed close to this temperature [7, 9].

The determine the low temperature (T < 5 K) behavior of HoMnO$_3$, we measure the low-T magnetization for magnetic fields between 5 and 20 kOe. The results are shown in Fig. 8. The most significant feature in M(T) on cooling is a sudden increase at a very low characteristic temperature $T_3(H)$. For example, at 6 kOe the magnetization jump by a factor of 1.7 occurs at $T_3$=2.1 K. $T_3(H)$ increases with H and the size of the magnetization step decreases until H reaches about 12 kOe. Above 12 kOe the magnetization step becomes negative and $T_3(H)$ decreases again (Fig. 8). The critical temperature of this anomaly defines a dome shaped phase region in the T-H phase diagram with a maximum $T_3$ of 3.3 K, shown in Fig. 9. This low temperature phase, tentatively called LT1 phase, is stable between approximately 5 kOe and 18 kOe, as indicated in Fig. 9 by the dotted lines extrapolating the phase boundaries to zero temperature. The sign change of the magnetization jump occurs right at the maximum of $T_3$ and it is in accordance with the Clausius-Clapeyron relation valid at first order phase transitions. Therefore, we conclude that the transition into the LT1 phase is of first order as it is also suggested by the step-like change of the magnetization at $T_3$. The thermodynamics of this transition is discussed in more detail in Section C.

The magnetization jump at $T_3$ results in a similarly sharp anomaly of $\chi'_{ac}(H)$, shown in Fig. 10. At the lowest temperature of 1.8 K the first narrow peak at 5 kOe appears at $H_3$ (as indicated in Fig. 10), the transition from the INT phase to the LT1 phase with increasing H. There is a second peak followed by a pronounced drop of $\chi_{ac}$ at 18 kOe (denoted by $H_1$ in Fig. 10). The latter anomaly shifts to higher magnetic field with increasing temperature and is identified as the extension of the $T_1(H)$ transition line from the high temperature part of the phase diagram as discussed above (see also Fig. 9).

At magnetic fields above 20 kOe, another anomaly in the c-axis magnetization appears indicative of a transition from the HT1 phase to a second low temperature phase, LT2. This anomaly is defined by a drop of the derivative, d(H/M)/dT, as shown in Fig. 4. The critical temperature for the HT1 – LT2 transition, $T_4(H)$, increases with the external field H and follows roughly the phase boundary between the AFM($B_2$) and AFM($A_1$) phases proposed in Ref. [6]. Fig. 5 and, in more detail, Fig. 9 summarize the phase diagram obtained from magnetic measurements. The most interesting feature is the existence of a tetracritical point at which four phases, INT, HT1, LT1, and LT2, meet at about 2 K and 18 kOe. Close to the tetracritical point the magnetic properties of HoMnO$_3$ change dramatically as, for example, indicated by steep changes and narrow peaks of the ac susceptibility (Fig. 10). The delicate coupling of magnetic orders (Ho-moments and Mn-spins) and the ferroelectric order obviously results in subtle changes of the magnetic properties and more phases with complex magnetic structures are possible in this part of the phase diagram. All the transitions across the phase boundaries denoted by $T_1(H)$, $T_2(H)$, and $T_3(H)$ are defined by sharp changes of the dc magnetization, the ac susceptibility, or the specific heat. The transition across $T_4$ as determined from the decrease of d(H/M)/dT in Fig. 4 did not appear to be very sharp and some uncertainty was involved in determining $T_4(H)$.

The phase diagrams in Figs. 5 and 9 were mainly determined from magnetic measurements with the field directed along the hexagonal c-axis. We have also conducted magnetic measurements with H // a. With the exception of the anomaly of M(T) at the zero field $T_2$=5.2 K (Fig. 2, inset) we have not detected any distinct anomaly similar to the c-axis data in crossing the various phase boundaries as discussed above. Since the Mn magnetic moments are assumed to stay in-plane (although in some magnetic space groups compatible with the crystalline group P6$_3$cm the Mn spins are allowed to have a c-axis component), the Ho magnetic moments may play a crucial role, probably coupled to the in-plane Mn spins, in determining the magnetic anomalies and phases observed in our experiments.

### B. Heat Capacity Measurements

The onset of magnetic order is also expected to result in anomalies of thermodynamic quantities such as the specific heat. The heat capacity, $C_P(T)$, of HoMnO$_3$ at zero magnetic field is shown in Fig. 11. The λ-type anomaly at $T_N$=76 K strongly suggests the second order nature of the AFM transition of the Mn$^{3+}$ magnetic moments. At the low temperature magnetic transition ($T_2$=5.2 K) $C_P(T)$ exhibits a sharp increase with decreasing T, enlarged in the upper inset of Fig. 11. Both anomalies are in qualitative agreement with recent measurements of polycrystalline samples.[3] In addition to these two pronounced anomalies we observe, for the first time, a narrow $C_P$-peak at the spin rotation transition temperature, $T_{SR}$. The details of this peak are displayed in the lower inset of Fig. 11. The position and the width of the peak are in perfect agreement with the c-axis magnetization anomaly (Fig. 3) and the peak of the dielectric constant observed by us before.[11] This is the first clear thermodynamic signature of the magnetic phase transition at 32.8 K. The calculated entropy change associated with this transition is ΔS=0.03 J/(mol K). Since the Mn-spin rotation does not contribute to the change of entropy, the ΔS observed is attributed to the AFM Ho-spin ordering at $T_{SR}$. However, the value of ΔS is far too small to account for a complete ordering of the Ho$^{3+}$ spin system. The origin of the small ΔS could be the partial ordering of only a fraction of the Ho moments or a high degree of short-range order that exists already above the transition temperature. The partial Ho ordering is compatible with the magnetization data discussed in the previous section. The paramagnetic contribution to the c-axis magnetization below $T_{SR}$ that amounts to about 70 % of the maximum effective moment of the Ho$^{3+}$ ion supports the assumption that the Ho moment order is not complete below $T_{SR}$. This result agrees well with the small sublattice magnetization of the $Ho^{3+}$ derived from neutron scattering data.[4] On the other hand, spin frustration effects in the quasi-triangular lattice of the Ho layers perpendicular to the c-axis may result in a high degree of short-range correlation well above the temperature below which long-range order eventually sets in. Both explanations have to be considered to understand the small entropy change at $T_{SR}$. It should be noted that the entropy change due to the AFM ordering of the $Mn^{3+}$ at $T_N$ is also much smaller than the maximum value of $Rln5$ expected for a complete order-disorder transition of the $Mn^{3+}$ moments with S=2 (we estimate an entropy change of about 10 to 15 % of $Rln5$, in agreement with Ref. [3]). The most plausible explanation is the existence of short-range order above $T_N$. The effects of geometric frustration on the short-range order and consequences for the entropy change at phase transitions have been discussed long time ago for the problem of charge order in the pyrochlore-type lattice of B-sites in the inverse spinel structure of magnetite. It was shown that the far smaller than expected change of entropy at the charge order – disorder transition (Verwey transition) in this compound can be explained by geometric frustration effects in the pyrochlore lattice.[12,13] We conclude that the observed small ∆S values at $T_N$ and at $T_{SR}$ of $HoMnO_3$ are of similar origin.

The λ-type anomaly at $T_N$ was not affected by an external magnetic field, which is in agreement with the specific heat data of polycrystalline samples.[3] This is an indication that the AFM spin order of the $Mn^{3+}$ ions is very rigid with respect to high magnetic fields. The Mn-spin rotation transition, however, strongly depends on the external field as shown in the phase diagram of Fig. 5. The field effect on $C_P(T)$ near 33 K is demonstrated in Fig. 12. The narrow peak of $C_P$ broadens and shifts to lower temperature with increasing H. The anomaly eventually disappears above 40 kOe. The $C_P$ peak temperature follows the phase transition lines $T_1(H)/T_2(H)$ determined from magnetic measurements in the previous section. A more accurate comparison and an assignment of the $C_P$ anomaly to either $T_1$ or $T_2$ in the high field region is not possible because of the broadening of the peak in high magnetic fields.

In the low temperature region the transition from HT2 to INT phase at $T_2$ is well defined by the sharp increase of $C_P$ that shifts to higher T with increasing field H, as shown in Fig. 13. The derived $T_2(H)$ data, plotted as green stars in Figs. 5 and 9, are in excellent agreement with the magnetically determined phase boundary. The most remarkable anomaly of the specific heat in the low temperature region is the extremely high and narrow peak that was detected at $T_3$, the transition into the LT1 phase. This peak emerges at about 5 kOe (at 1.8 K) and rapidly increases up to 12 kOe (at 3.3 K, the maximum value of $T_3$). With further increasing H the $C_P$ peak decreases again and vanishes above 18 kOe. The position of the peak is in perfect agreement with the $T_3(H)$ data determined in the previous section (Fig. 9). The step-like change of the magnetization and the sharp anomaly of $C_P$ at $T_3(H)$ are characteristics of a first order phase transition. The entropy change associated with this transition at 12 kOe/3.3 K is about 0.5 J/mol K, an order of magnitude larger than that at $T_{SR}$. We propose that this phase transition into the LT1 phase involves a major ordering of the $Ho^{3+}$ magnetic moments.

The heat capacity above 25 kOe did not show very pronounced or sharp anomalies in the low-T region (T < 10 K) of the phase diagram. In particular, the magnetic phase change at $T_4$ from the HT1 to the LT2 phase appeared without distinct thermodynamic anomalies. This is in line with the broad change of the magnetization (or the derivative d(H/M)/dT, cf. Fig. 4) at $T_4$ as already discussed in the previous section.

### C. The Unified Magnetic Phase Diagram

The magnetic and thermodynamic data demonstrate that the field-induced intermediate phase (INT) between two high temperature phases (HT1 and HT2) previously suggested [11] extends to low temperature at low field and define a novel and yet complex unified phase diagram of $HoMnO_3$. Four phases, INT, HT1, LT1, and LT2 are unambiguously identified as summarized in Figs. 5 and 9. An extremely unusual tetracritical point at $T^* = 2$ K and $H^* = 18$ kOe is clearly evident. The phase boundary between the HT1 and LT2 phases is in good agreement with that proposed recently based on optical experiments.[6] However, the existence of the INT phase and the details of the low temperature phase diagram with the dome shaped stability region of the LT1 phase (Fig. 14) are in striking difference to the phase diagram of Ref. [6]. The enhanced ac magnetic susceptibility observed in this phase is similar to the enhanced dielectric constant reported earlier[11], indicating an unusual coupling between magnetic and dielectric orders.

The first order transition line $T_3(H)$ separating the INT and LT1 phases in Fig. 9 is well characterized with respect to its magnetic as well as thermodynamic anomalies. At a first order magnetic phase transition the abrupt changes of M(T) and S(T) are correlated via the Clausius-Clapeyron relation, $dT_c/dB = -\Delta M/\Delta S$, where $B = \mu_0 (H+M_v)$ is the magnetic induction, $M_v$ is the volume magnetization, and $\Delta M$ and $\Delta S$ are the magnetization jump and entropy change across the transition, respectively. To verify the thermodynamic consistency by employing the Clausius-Clapeyron relation the phase boundary $T_3(B)$ has to be plotted as a function of B, calculated from the external field H and the volume magnetization $M_v$. Fig. 14 shows the phase boundary $T_3(B)$ with the critical temperatures estimated from dc magnetization, ac susceptibility, and specific heat experiments. The line is a fit of a fourth order polynomial to the data. At the apex of the dome shaped $T_3(B)$ near 1.4 Tesla the derivative $dT_3/dB$ changes sign which infers that also $\Delta M$ has to change sign ($\Delta S$ is positive). This sign change was, in fact, observed in our magnetic measurements (see discussion in the previous section). A more careful evaluation of the ratio $\Delta M/\Delta S$ extracted from our magnetic and heat capacity data allows us to prove the consistency of the data with the Clausius-Clapeyron equation. The derivative, $dT_3/dB$, is shown in the inset of Fig. 14 (line). The values of $-\Delta M/\Delta S$ plotted as stars in the same figure are in perfect agreement with the slope $dT_3/dB$. This confirms unambiguously the first order nature of the transition into the LT1 phase.

The assignment of magnetic symmetries and the details of the magnetic structure in the various phases cannot be derived from our current experimental results. The following discussion will be tentative and has to rely also on previously published data. As to the high temperature region (T>5 K) of the phase diagram, neutron scattering experiments on polycrystalline samples [3,4] as well as optical experiments on single crystals [5,6] indicate that the magnetic space group in our HT1 phase below $T_N$ is P$\underline{6_3}$cm. In this symmetry the $Mn^{3+}$ moments are aligned in-plane perpendicular to the hexagonal a-axis and neighboring moments form a 120° angle. No c-axis component of the moment is allowed. For a detailed presentation of the spin orders in the different magnetic symmetry classes we refer the reader to the discussion in Ref. [6]. At $T_{SR}$ the $Mn^{3+}$ moments rotate in-plane by 90° so that they become aligned with the a-axis. The resulting space group is P$\underline{6}_3$c$\underline{m}$, the same as in YMnO$_3$ below the Néel transition temperature. An out-of-plane component of the $Mn^{3+}$ moment is not forbidden in this symmetry, but so far there is no experimental evidence for it. The transition from the HT1 to the HT2 phase proceeds via the INT phase the symmetry of which is still unknown. It appears conceivable that the transition from P$\underline{6}_3$cm to P$\underline{6}_3$c$\underline{m}$ involves the lower P$\underline{6}_3$ symmetry with an alignment of the Mn-moments at an intermediate angle between 90° and 0° with the a-axis. This symmetry is compatible with neutron scattering data in a broader temperature range (33 K < T < 42 K at zero magnetic field) [4] and with optical experiments close to $T_{SR}$.[5] According to our phase diagram (Fig. 5) the INT phase (P$\underline{6}_3$) at H=0 does not extend over a larger temperature range, the transition from P$\underline{6}_3$cm to P$\underline{6}_3$c$\underline{m}$ symmetry happens within a few tenths of a Kelvin. The broader stability range reported in Ref. [4] at H=0 is obviously a consequence of the polycrystalline nature of the sample. Only at finite magnetic fields will the INT phase region spread over a wider temperature range between $T_1(H)$ and $T_2(H)$ as shown in Fig. 5.

The low-temperature (T<5 K) part of the phase diagram (Fig. 9) is substantially different from previous reports.[4,6] The INT phase appears to extend to the lowest temperatures we have examined for magnetic fields below 5 kOe. This explains some recent experimental results including the observation of a distinct step of the dielectric constant at $T_2(H=0)$=5.2 K, as discussed in Section III A. It should be noted that recent reports disagree about the symmetry assignment of the H=0 low temperature phase. Neutron scattering experiments [4] suggested the low temperature phase be of the P6$_3$cm symmetry class whereas optical measurements favored P6$_3$c$\underline{m}$.[6] At high magnetic field the LT2 phase boundary, $T_4(H)$, is in good agreement with the AFM(A$_1$) phase of Ref. [6]. The stability region of the LT1 phase with its dome shaped phase boundary $T_3(H)$, however, is in contrast to recent reports.[6] $T_3(H)$ is very well defined by the sharp magnetization anomaly and the narrow peak of the heat capacity. It provides clear evidence that the low-T phase diagram of HoMnO$_3$ is far more complex than previously assumed. In particular, the four phases INT, HT1, LT1, and LT2 (but not the HT2 phase) join at the tetracritical point (at 2 K, 18 kOe). Following Ref. [6], the P6$_3$c$\underline{m}$ and P6$_3$cm symmetries could be assigned to the low-T phases LT1 and LT2, respectively.

The magnetic structure of the various phases including the AFM order of the Ho$^{3+}$ moments have been explored by neutron scattering experiments of powdered samples. Our thermodynamic data are consistent with some recent results. For example, the small sublattice magnetization of the Ho-ions measured right below $T_{SR}$ [4] is in accordance with the small entropy change observed at this temperature. However, the exploration of the magnetic orders of the different species (Mn$^{3+}$ and Ho$^{3+}$) and their mutual correlation in the whole T-H phase diagram requires far more experimental work. We propose to conduct neutron scattering experiments using single crystals of HoMnO$_3$. Work is underway with our success in growing large single crystals of HoMnO$_3$ by the floating zone technique.

## IV. SUMMARY

A novel magnetic phase diagram with a tetracritical point has been established in the hexagonal $HoMnO_3$ by our magnetic and heat capacity experiments. Two new field-induced phases have been discovered. Together with those previously known, there are at least five different phases existing below the Néel temperature and for magnetic fields up to 50 kOe. The phases are well characterized by their magnetic properties and the phase boundaries are defined by distinct anomalies in their magnetization, ac susceptibility, and specific heat. In particular, an intermediate phase between the P6$_3$cm and P6$_3$cm magnetic structures was unambiguously shown to exist below 33 K (the Mn-spin rotation transition temperature) and at magnetic fields H>0 in accordance with the conclusions of recent dielectric measurements. This intermediate phase extends to lower temperatures. The most interesting feature in the low temperature region of the T-H phase diagram is a dome shaped first order phase transition line joining other phase boundaries at a tetracritical point at 2 K and 18 kOe. We prove that the magnetic and thermodynamic data along the line of first order transitions are consistent with the magnetic analogue of the Clausius-Clapeyron equation. Our detailed exploration of the phase diagram shows a far more complex phase structure than previously assumed.

## ACKNOWLEDGMENTS

The authors would like to thank A. P. Litvinchuk and M. Iliev for stimulating discussions. This work is supported in part by NSF Grant No. DMR-9804325, the T.L.L. Temple Foundation, the John J. and Rebecca Moores Endowment, and the State of Texas through the TCSAM at the University of Houston and at Lawrence Berkeley Laboratory by the Director, Office of Energy Research, Office of Basic Energy Sciences, Division of Materials Sciences of the U.S. Department of Energy under Contract No. DE-AC03-76SF00098. The work of M. M. G. is supported by the Bulgarian Science Fund, grant No F 1207.

**References**

[1]  Z. J. Huang, Y. Cao, Y. Y. Sun, Y. Y. Xue, and C. W. Chu, Phys. Rev. B **56**, 2623 (1997).
[2]  A. Munoz, J. A. Alonso, M. J. Martinez-Lope, M. T. Casais, J. L. Martinez, and M. T. Fernandez-Diaz, Phys. Rev. B **62**, 9498 (2000).
[3]  A. Munoz, J. A. Alonso, M. J. Martinez-Lope, M. T. Casais, J. L. Martinez, and M. T. Fernandez-Diaz, Chem. Mater. **13**, 1497 (2001).
[4]  Th. Lonkai, D. Hohlwein, J. Ihringer, and W. Prandl, Appl. Phys. A **74**, 843 (2002).
[5]  M. Fiebig, D. Fröhlich, K. Kohn, St. Leute, Th. Lottermoser, V. V. Pavlov, and R. V. Pisarev, Phys. Rev. Letters **84**, 5620 (2000).
[6]  M. Fiebig, C. Degenhardt, and R. V. Pisarev, J. Appl. Phys. **91**, 8867 (2002).
[7]  N. Iwata and K. Kohn, J. Phys. Soc. Jpn. **67**, 3318 (1998).
[8]  T. Katsufuji, S. Mori, M. Masaki, Y. Moritomo, N. Yamamoto, and H. Takagi, Phys. Rev. B **64**, 104419 (2001).
[9]  H. Sugie, N. Iwata, K. Kohn, J. Phys. Soc. Jpn. **71**, 1558 (2002).
[10] R. R. Birss, Symmetry and Magnetism (North-Holland, Amsterdam, 1966).
[11] B. Lorenz, A. P. Litvinchuk, M. M. Gospodinov, and C. W. Chu, Phys. Rev. Letters **92**, 087204 (2004).
[12] P. W. Anderson, Phys. Rev. **102**, 1008 (1956).
[13] D. Ihle and B. Lorenz, Phys. Stat. Sol. (b) 63, 599 (1974).

Figure Captions

Fig. 1. Single crystal of HoMnO$_3$ used for magnetic measurements. The largest linear dimension is 2.6 mm and the thickness of this crystal is 0.26 mm.

Fig. 2. Inverse magnetization (after demagnetization correction) of HoMnO$_3$, measured along a- (squares) and c-axes (circles). The lines indicate the Curie-Weiss high-temperature extrapolation. Inset: Low-temperature region with the anomalies at 5.2 K.

Fig. 3. Inverse c-axis magnetization (open symbols) below 40 K and its derivative (closed symbols). The two critical temperatures, $T_{SR}$ and $T_2$, are marked by the narrow peaks of d(H/M)/dT.

Fig. 4. (Color online) The derivative of the inverse magnetization at different external magnetic fields:
0.1 (A), 25 (B), 33 (C), 35 (D), 38 (E), and 45 kOe (F). The curves (A) to (E) are offset by different constants for higher clarity. If all curves are plotted at the same scale they coincide at their high-temperature values, as shown in Fig. 6. The INT phase is stable between pairs of $T_1$ and $T_2$ as defined for curve C (33 kOe).

Fig. 5. (Color online) Phase diagram of HoMnO$_3$ below the Néel temperature. Different phase boundaries are indicated by $T_1$ to $T_4$ as determined from anomalies of the dc magnetization (open circles), ac susceptibility (closed circles) and heat capacity (stars).

Fig. 6. (Color online) The derivative d(H/M)/dT at different magnetic fields between 0.1 and 45 kOe plotted on the same scale. The dotted line defines the temperature dependence of d(H/M)/dT in the intermediate (INT) phase.

Fig. 7. (Color online) Isothermal ac susceptibility as function of the external field for temperatures above 7 K.

Fig. 8. (Color online) Low temperature magnetization of HoMnO$_3$ measured in different magnetic fields. The applied fields from top to bottom curves are: 20, 18, 17.5, 17, 16.5, 16, 14, 12, 11, 10, 9, 8, 6, and 5 kOe. The two anomalies defining $T_2$(H) (right) and $T_3$(H) (left) are shown by the dotted lines.

Fig. 9. (Color online) Low-temperature part of the phase diagram of HoMnO$_3$. Four phases coexist at the tetracritical point (2 K, 18 kOe). The symbols are chosen as in Fig. 5.

Fig. 10. (Color online) Field dependence of the ac susceptibility at low temperatures (T<7 K). The sharp peak at the low-field end defines the phase boundary $T_3$(H) and the drop at the high-field side identifies $T_1$(H), as shown in Fig. 9.

Fig. 11. Specific heat of $HoMnO_3$ at zero magnetic field. The λ-type anomaly at $T_N$ is followed by two additional peaks at $T_{SR}$ and $T_2$, enlarged in the lower right and upper left insets, respectively.

Fig. 12. (Color online) Field dependence of the heat capacity anomaly close to $T_{SR}$.

Fig. 13. (Color online) Low-temperature heat capacity for different magnetic fields up to 20 kOe. The high-temperature edge defines the phase boundary $T_2(H)$ and the sharp peak at low temperatures indicates the first order phase transition across $T_3(H)$.

Fig. 14. (Color online) Phase boundary $T_3$ defining the stability region of the LT1 phase as a function of magnetic induction B. The inset proves the thermodynamic consistency of magnetic and heat capacity data (the magnetic analogue of the Clausius-Clapeyron equation valid at first order phase boundaries is fulfilled).

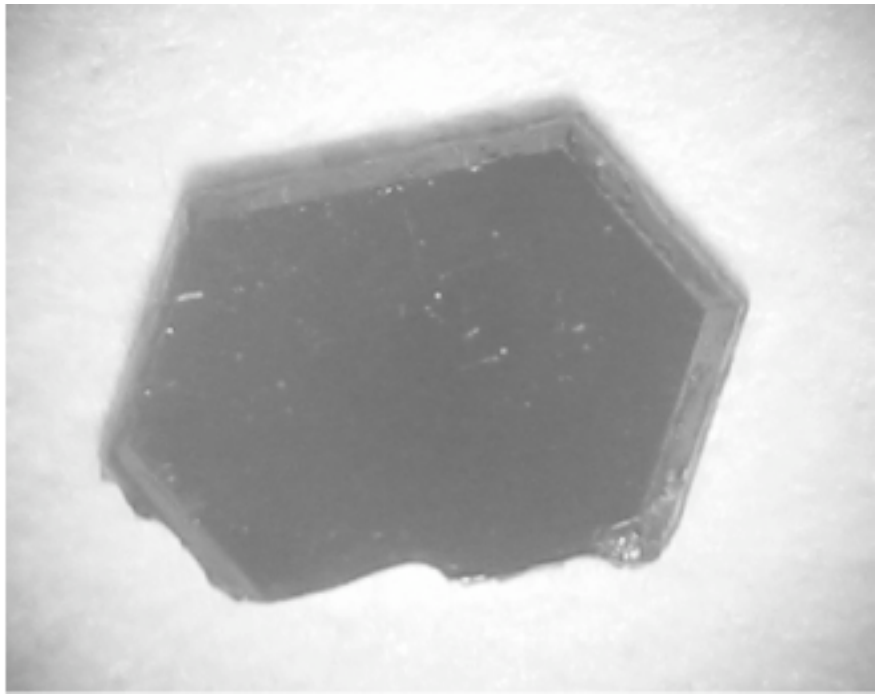

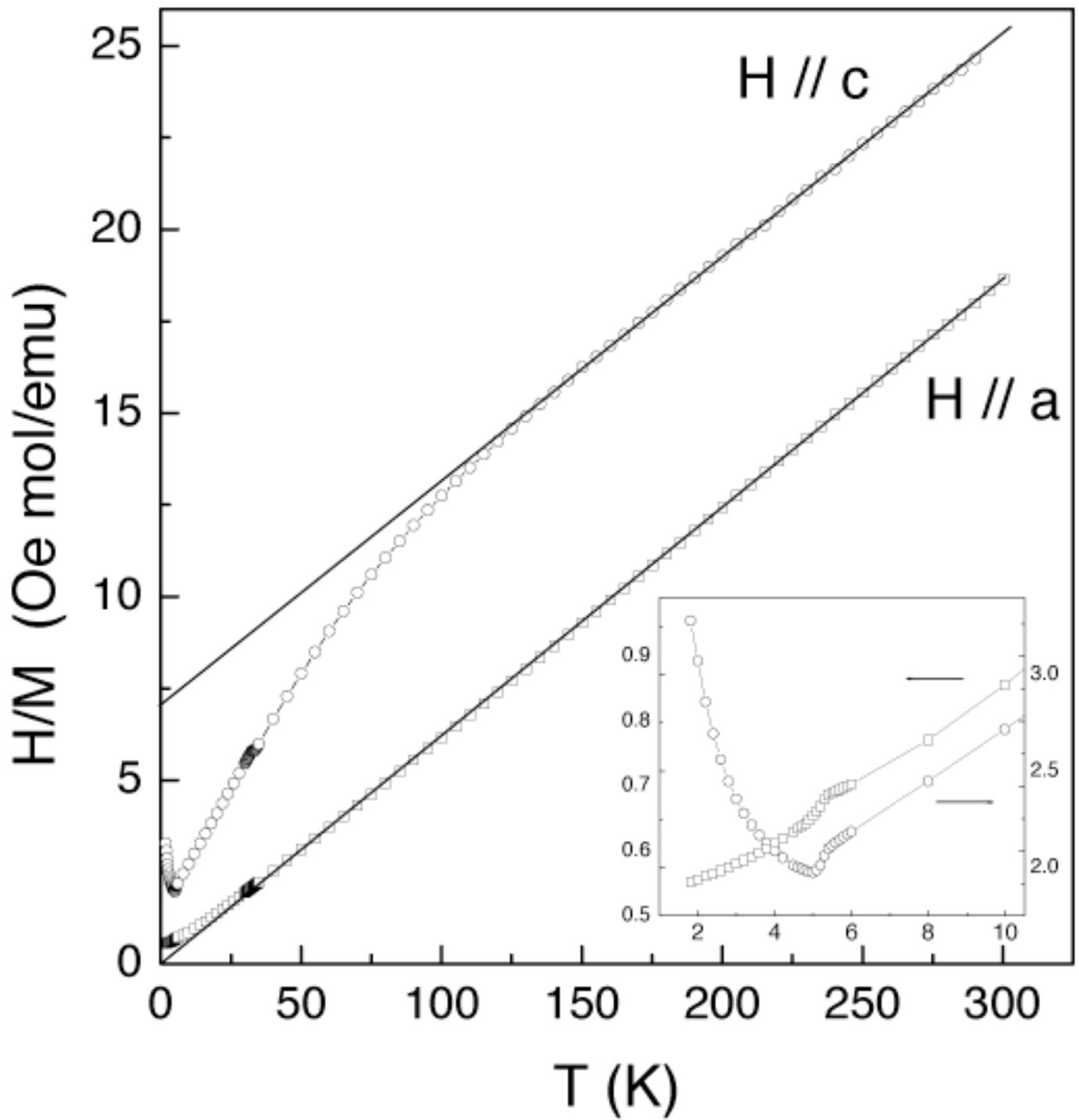

Fig. 2
B. Lorenz et al.

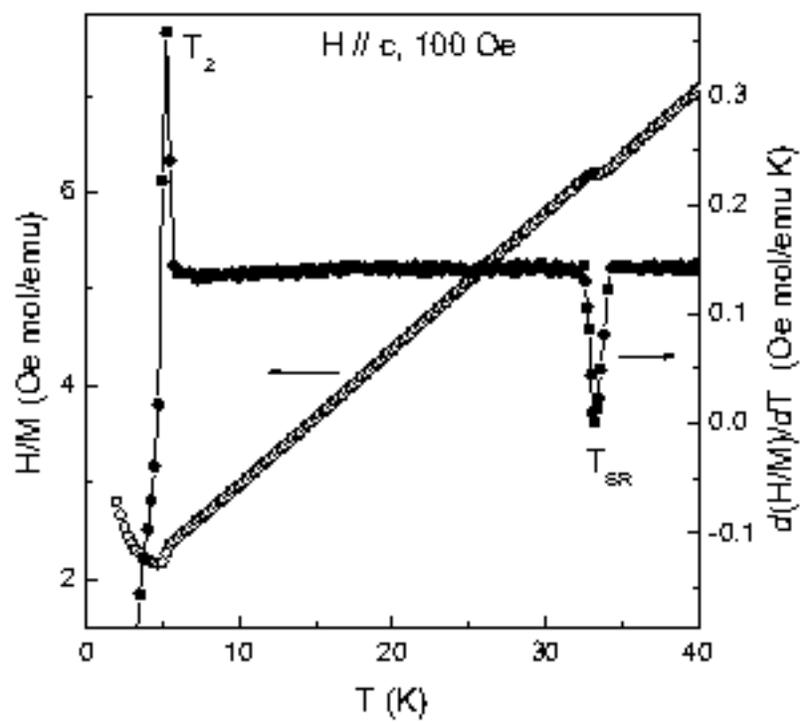

Fig. 3
B. Lorenz et al.

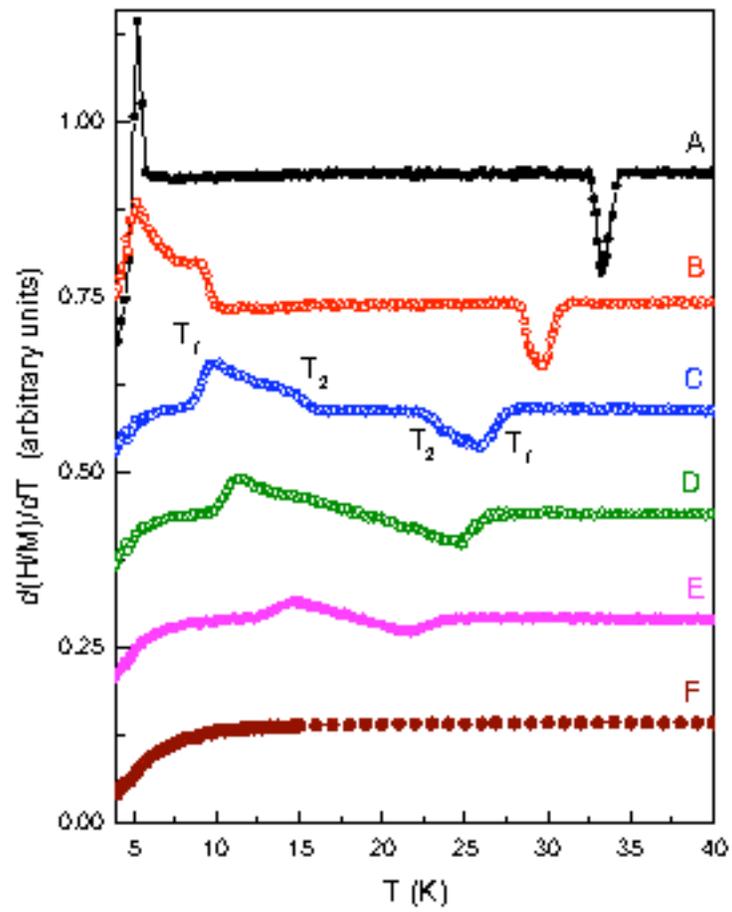

Fig. 4
B. Lorenz et al.

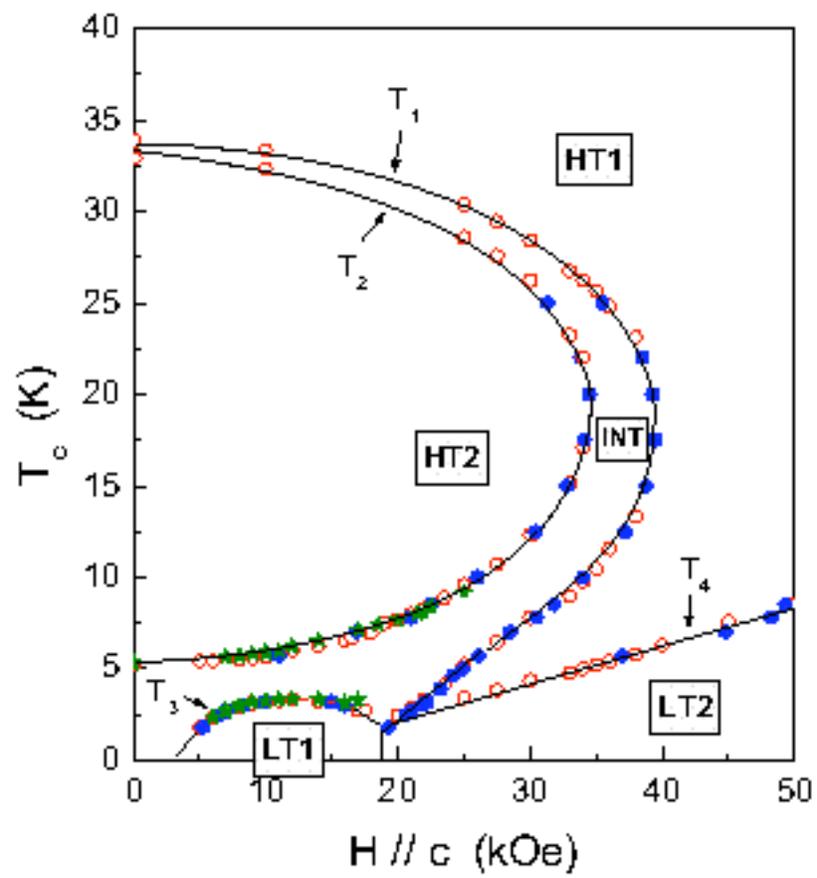

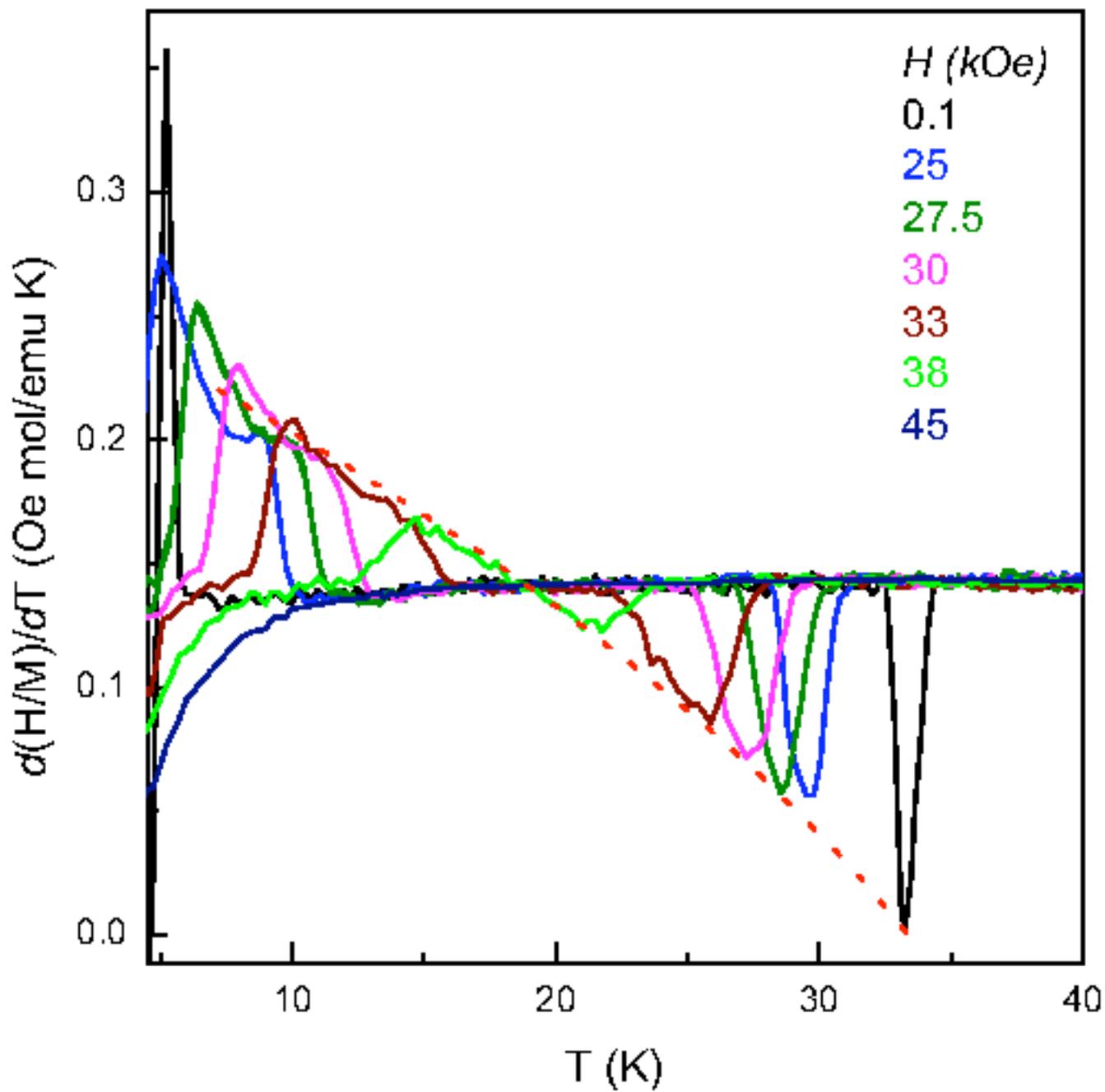

Fig. 6
B. Lorenz et al.

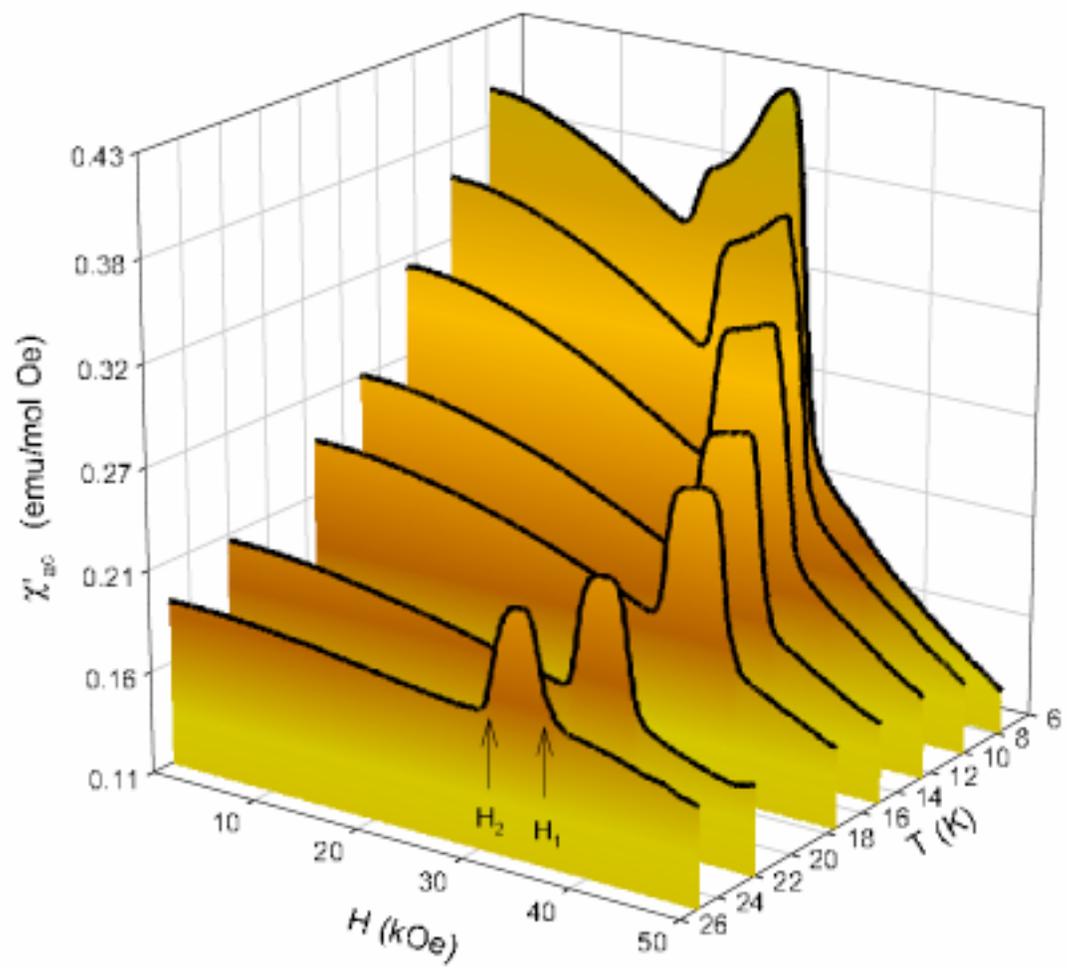

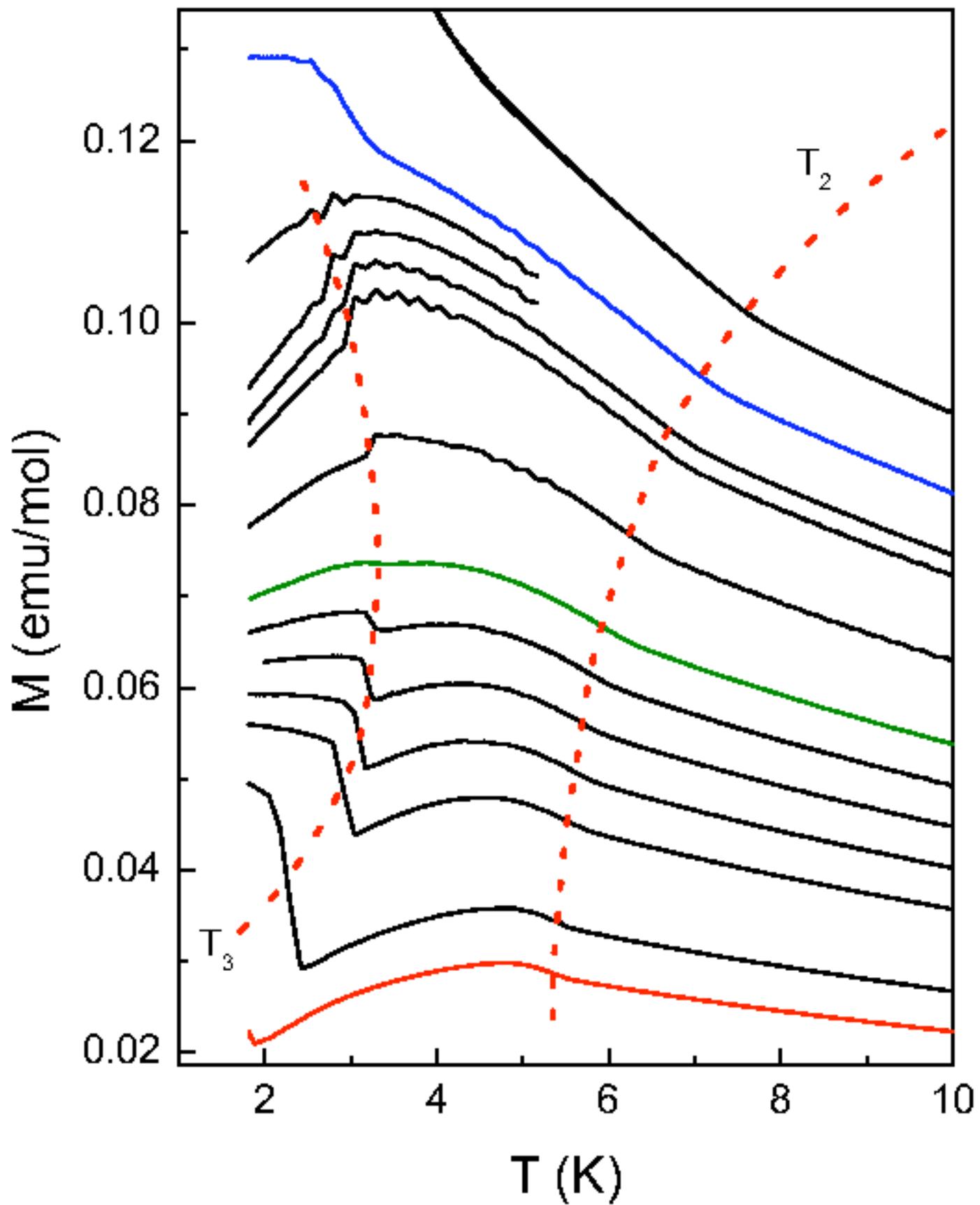

Fig. 8
B. Lorenz et al.

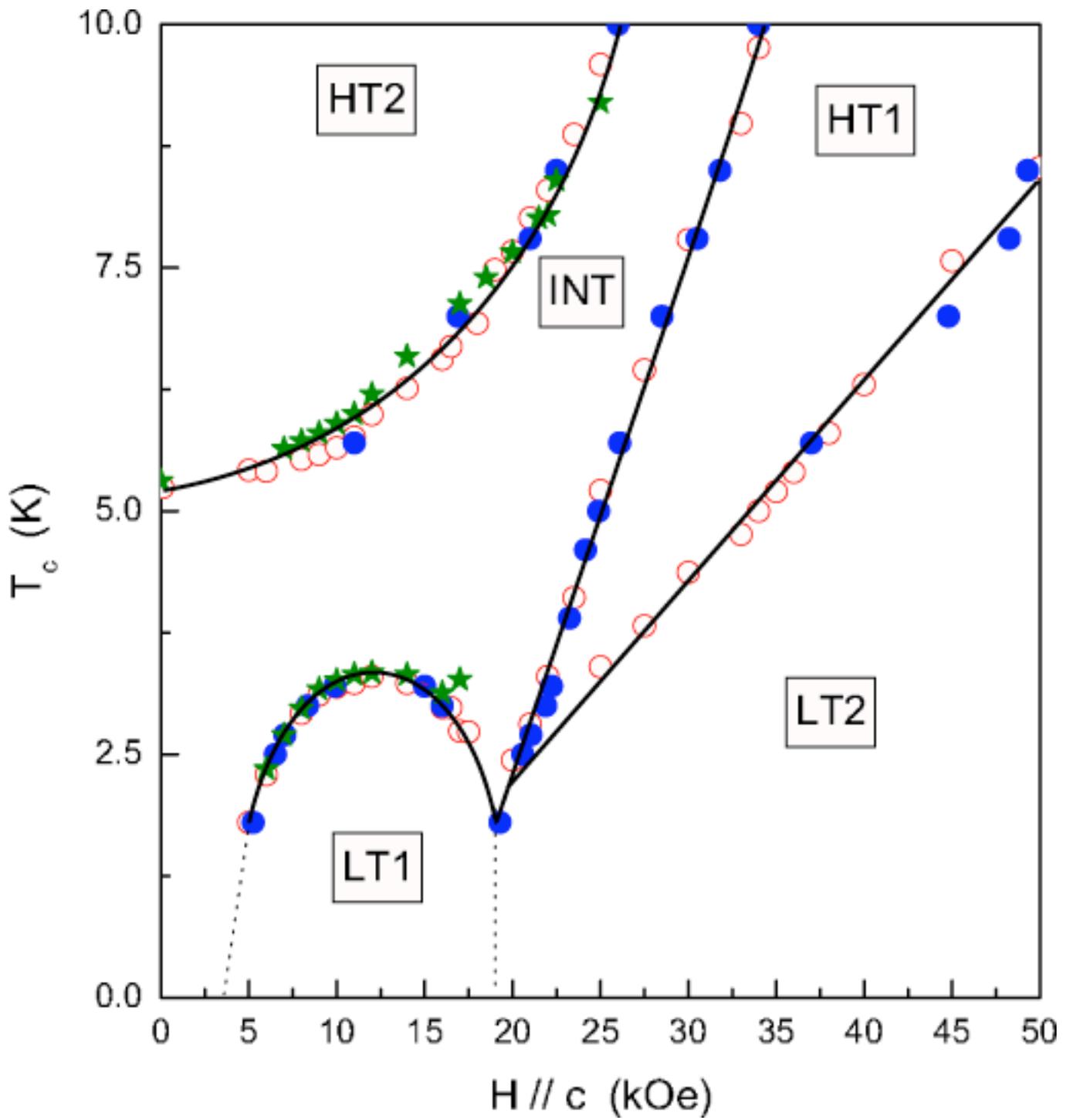

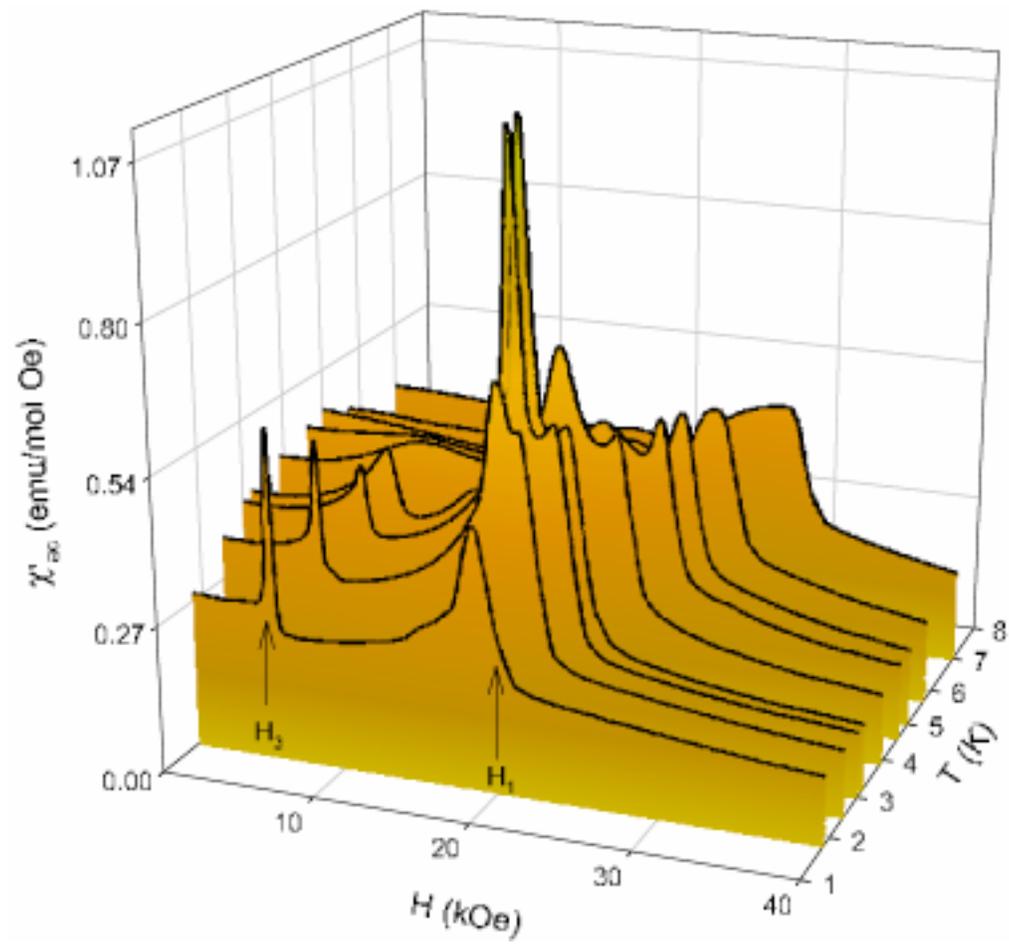

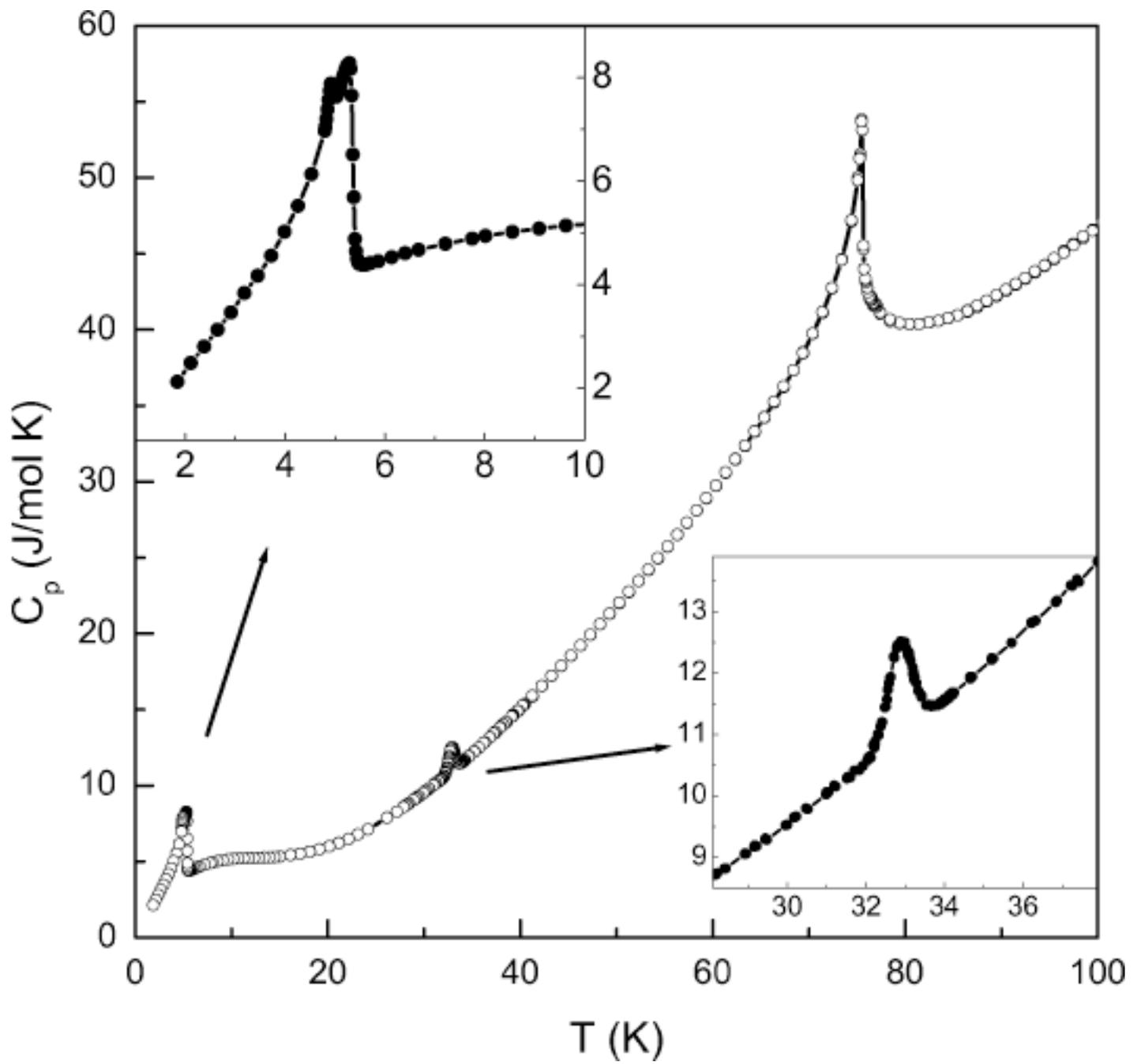

Fig. 11
B. Lorenz et al.

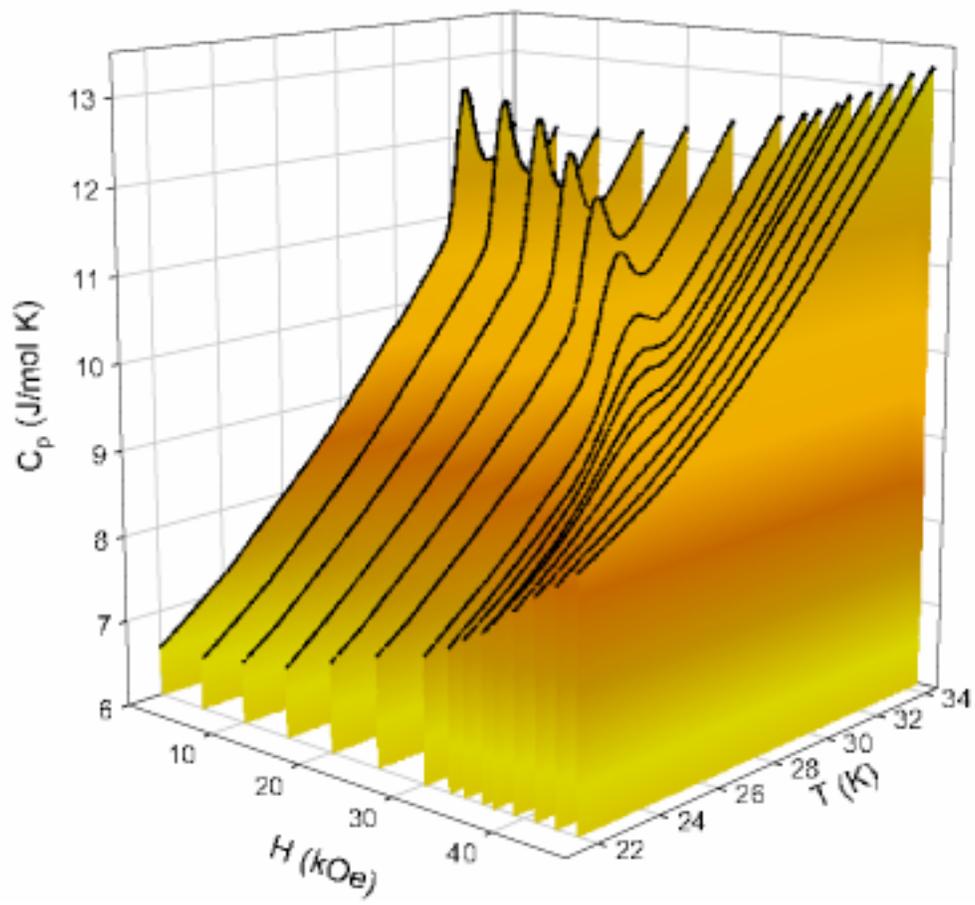

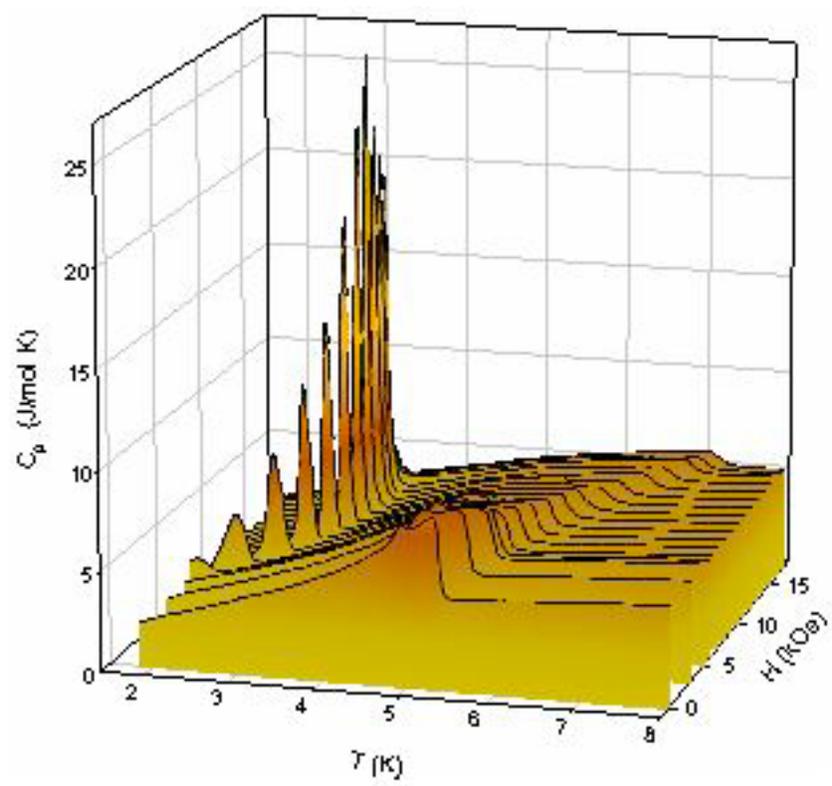

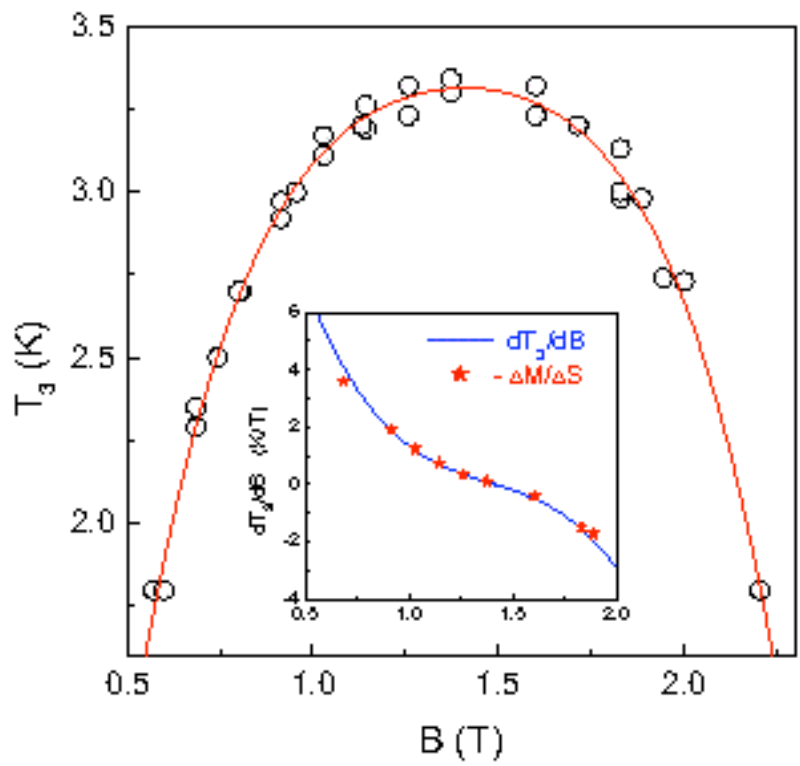

Fig. 14
B. Lorenz et al.